\documentclass[preprint,12pt,authoryear]{elsarticle}

\usepackage{amssymb}
\usepackage{amsmath}
\usepackage{float}
\journal{arXiv.org}

\begin{document}

\begin{frontmatter}



\title{New Stress‐dependent Elastic Wave Velocity Models for Reservoir Rocks with Applications}


\author[inst1,inst2]{Rong Zhao}

\affiliation[inst1]{organization={Sinopec Key Laboratory of Ultra-Deep Well Drilling Engineering Technology}, %Department and Organization
            city={Beijing},
            postcode={102206}, 
            country={China}}
\affiliation[inst2]{organization={SINOPEC Research Institute of Petroleum Engineering Co. Ltd}, %Department and Organization
            city={Beijing},
            postcode={102206}, 
            country={China}}
%Sinopec Key Laboratory of Ultra-Deep Well Drilling Engineering Technology, Beijing 102206, China
%SINOPEC Research Institute of Petroleum Engineering 
%Co. Ltd, Beijing 102206, China

\author[inst3]{Chunguang Li}

\affiliation[inst3]{organization={State Key Laboratory of Geomechanics and Geotechnical Engineering, Institute of Rock and Soil Mechanics, Chinese Academy of Sciences},%Department and Organization
            city={Wuhan},
            postcode={430071}, 
            country={China}}

\begin{abstract}
This study presents new elastic velocity-effective stress laws for reservoir rocks. These models are grounded in previously established correlations between elastic modulus and porosity, which incorporate critical porosity. The accuracy of the models is validated against wave velocities from 38 core samples, yielding coefficients of determination ($\mathrm{R}^2$) of 0.9994 for compressional wave and 0.9985 for shear wave. A sensitivity analysis reveals that the maximum uncertainties for compressional and shear waves are less than $\pm$5.5\% and $\pm$7.5\%, respectively. To demonstrate the applicability of the proposed models, a case study was conducted on three wells in the Northern Carnarvon Basin, where the new elastic wave velocity-effective stress laws produced reliable predictions for velocity logs in the studied formations. The relationships reported herein may prove beneficial for hydrocarbon exploration, production, and ensuring drilling safety in both unconventional and conventional fields.
\end{abstract}




\end{frontmatter}


\section{Introduction}
Understanding elastic velocities in porous media is of considerable interest across various research fields, including rock mechanics, geological engineering, geophysics, and petroleum exploration \citep{Khaksar1999}. It is well established that elastic velocities are influenced by several factors, including porosity, pore-filling minerals, internal and external pressures, pore geometry, and pore fluid saturation \citep{Wei2022, Ba2024}. Among these factors, effective stress is particularly critical, especially in abnormally high-pressure formations, shallow gas layers, and unconsolidated sandstone formations \citep{Ge2001}. Consequently, investigating the effects of effective stress on wave velocities is of significant importance.

%目前，许多学者通过试验测定了不同岩石在不同有效应力下的弹性波速，得到基本结论为，波速随有效应力增加而增加，并趋于常熟。然而，弹性波速和有效应力的理论关系表达式还比较缺乏，大多数以经验为主，并且模型及模型参数的物理意义还有待深化。

Currently, numerous researchers have measured the elastic wave velocities of various rock types under different effective stresses \citep{Birch1960, Nur1969, Tosaya1982, Coyner1984, Wepfer1991, Freund1992, Blangy1992, Blangy1993}. The prevailing conclusion is that wave velocity increases with effective stress and tends to stabilize at a constant value. However, theoretical expressions that describe the relationship between elastic wave velocities and effective stress are still lacking, as most studies have relied on empirical formulas \citep{Freund1992, He2021}. Furthermore, the physical significance of the models and their parameters necessitates further clarification \citep{Ge2001}.

The outline of this paper is as follows. First, we establish new models for the longitudinal and shear waves in reservoir rocks in Section \ref{sec2}. This is followed by validation and a case study presented in Section \ref{sec3}. Finally, we conclude the paper in Section \ref{sec4}.

\section{Elastic Wave Velocity-Effective Stress Laws} \label{sec2}

\subsection{Porosity dependence of velocity of elastic waves}
The porosity-dependent elastic velocities of rocks can be expressed in terms of the zero-porosity elastic velocities and porosity \citep{zhao2021}, as follows:
\begin{equation} \label{eq1}
\begin{cases}
   v_{\mathrm{l}}=v_{\mathrm{lm}}\sqrt{(1-c_{\mathrm{l}}\phi)(1-\phi)}, \\
   v_{\mathrm{s}}=v_{\mathrm{sm}}\sqrt{(1-c_{\mathrm{s}}\phi)(1-\phi)},
\end{cases}
\end{equation}
where $c_{\mathrm{l}}=\frac{3\left(9K_{\mathrm{m}}^{2}-4K_{\mathrm{m}}G_{\mathrm{m}}+16G_{\mathrm{m}}^{2}\right)}{4G_{\mathrm{m}}\left(9K_{\mathrm{m}}+8G_{\mathrm{m}}\right)}$, $c_{\mathrm{s}}=\frac{6K_{\mathrm{m}}+12G_{\mathrm{m}}}{9K_{\mathrm{m}}+8G_{\mathrm{m}}}$, $v_\mathrm{lm}$ and $v_\mathrm{sm}$
represent the velocities of longitudinal and shear waves in the matrix, respectively. $K_\mathrm{m}$ and $G_\mathrm{m}$ denote zero-porosity bulk and shear moduli, respectively.

\subsection{Effective stress and porosity relationship}
It is widely accepted that formation porosity and effective
stress have the following relationship (e.g., \citet{Schoen2011}; \citet{Zhang2011}; \citet{Zhang2013}; ), expressed as:
\begin{equation} \label{eq2}
    \phi=\phi_0e^{-c\sigma_\mathrm{e}},
\end{equation}
where $\phi_0$ is the initial porosity, $c$ is the stress compaction constant, and $\sigma_\mathrm{e}$ is effective stress, defined as the difference between the
overburden pressure and the pore pressure.

\subsection{Proposed laws}
Substitution of Eq.(\ref{eq2}) in Eq.(\ref{eq1}) gives the result
\begin{equation} \label{eq3}
\begin{cases}
   v_{\mathrm{l}}=v_{\mathrm{lm}}\sqrt{(1-c_{\mathrm{l}}\phi_{0}\mathrm{e}^{-c\sigma_{\mathrm{e}}})(1-\phi_{0}\mathrm{e}^{-c\sigma_{\mathrm{e})}}},\\
   v_{\mathrm{s}}=v_{\mathrm{sm}}\sqrt{(1-c_{\mathrm{s}}\phi_{0}\mathrm{e}^{-c\sigma_{\mathrm{e}}})(1-\phi_{0}\mathrm{e}^{-c\sigma_{\mathrm{e}})}}.
\end{cases}
\end{equation}

If there is no pore pressure, the confining pressure equals the effective pressure. Therefore, Eq.(\ref{eq3}) can be expressed as follows:
\begin{equation} \label{eq4}
\begin{cases}
   v_{\mathrm{l}}=v_{\mathrm{lm}}\sqrt{(1-c_{\mathrm{l}}\phi_{0}\mathrm{e}^{-c\sigma_{\mathrm{e}}})(1-\phi_{0}\mathrm{e}^{-c\sigma_{\mathrm{c})}}},\\
   v_{\mathrm{s}}=v_{\mathrm{sm}}\sqrt{(1-c_{\mathrm{s}}\phi_{0}\mathrm{e}^{-c\sigma_{\mathrm{e}}})(1-\phi_{0}\mathrm{e}^{-c\sigma_{\mathrm{c}})}}.
\end{cases}
\end{equation}

\section{Results} \label{sec3}

In this study, experimental data points and velocity logs published in the literature were fitted to Eq.(\ref{eq3}) for validation purposes. The goodness of fit was assessed using the coefficient of determination, $\mathrm{R}^2$. The resulting optimal coefficients from this fitting process serve as reference values for sensitivity analysis. Additionally, the relative change in velocity was employed as a statistical measure to evaluate the robustness of the proposed models across different effective pressures.

\subsection{Validation from core samples data}

\begin{figure}[H]
\centering
\includegraphics[width=0.65\textwidth]{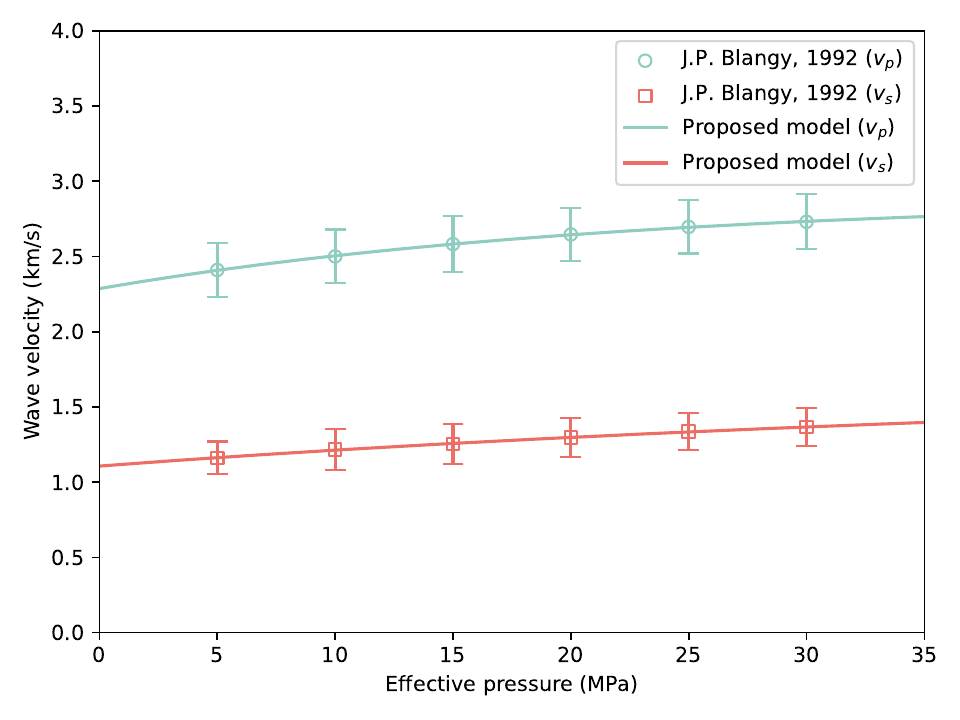}
\caption{Compressional and shear velocities as a function of effective pressure. Each data point marked with circles or squares is an average of over 38 core samples for effective pressure. Error bars represent standard deviations. The corresponding solid lines represent the fitting curves.  (Color online)}
\label{fig1}
\end{figure}

\citet{Blangy1992,Blangy1993} investigated 38 core samples from an
80 m cored section in the Sognefjord sands. Compressional and shear wave velocities in fully water-saturated samples, with
pore pressure equal to 15 MPa, under confining pressures of 20, 25, 30, 35, 40, and
45 MPa, yielding effective pressures of 5, 10, 15, 20, 25, and 30 MPa, respectively.  Fig.\ref{fig1} represents the measured longitudinal and transverse velocities as a function of effective pressure, respectively. The $\mathrm{R}^2$
 of measured and predicted values are 0.9994 and 0.9985 for longitudinal and transverse velocities, respectively.

\begin{figure}[H]
\centering
\includegraphics[width=0.75\textwidth]{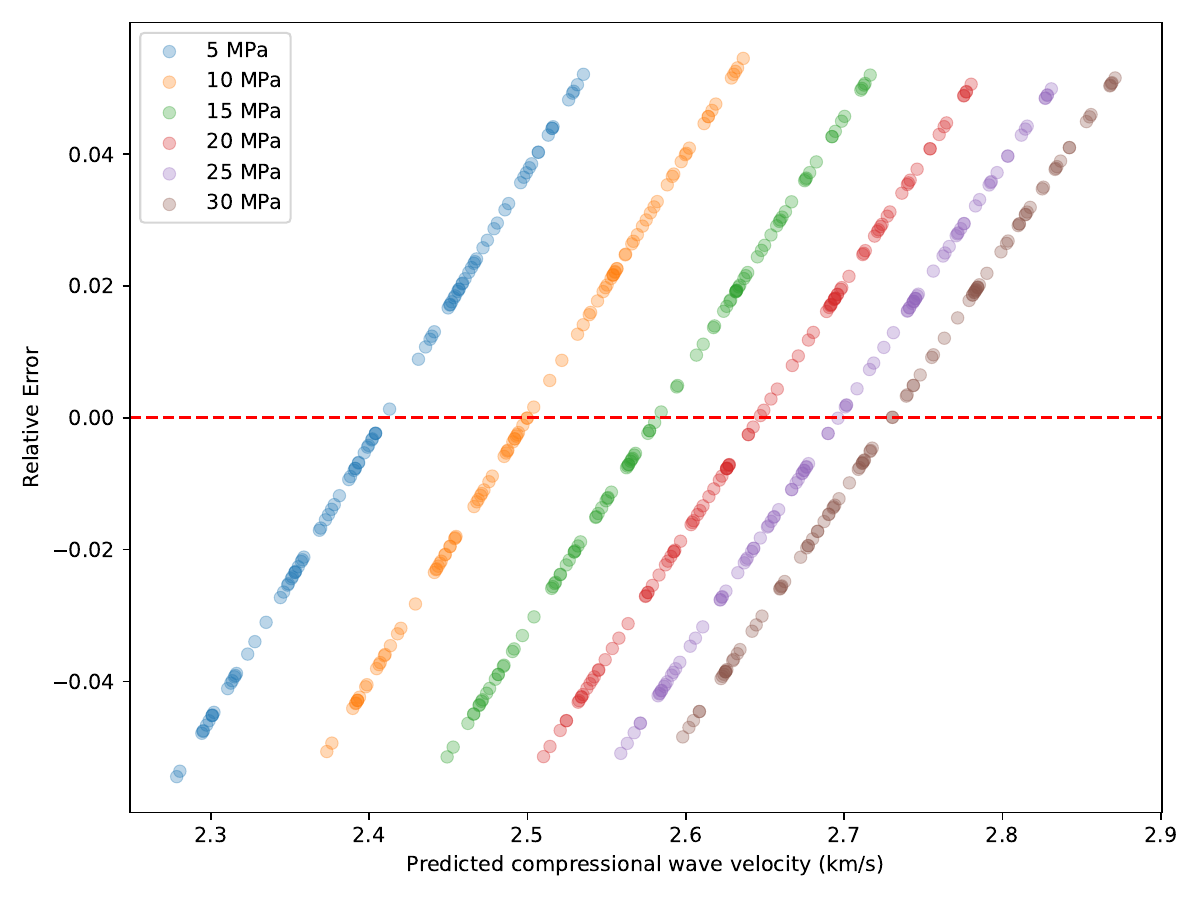}
\caption{Sensitivity analysis on the uncertainty of longitudinal velocity for each effective pressure. (Color online)}
\label{fig2}
\end{figure}

\begin{figure}[H]
\centering
\includegraphics[width=0.75\textwidth]{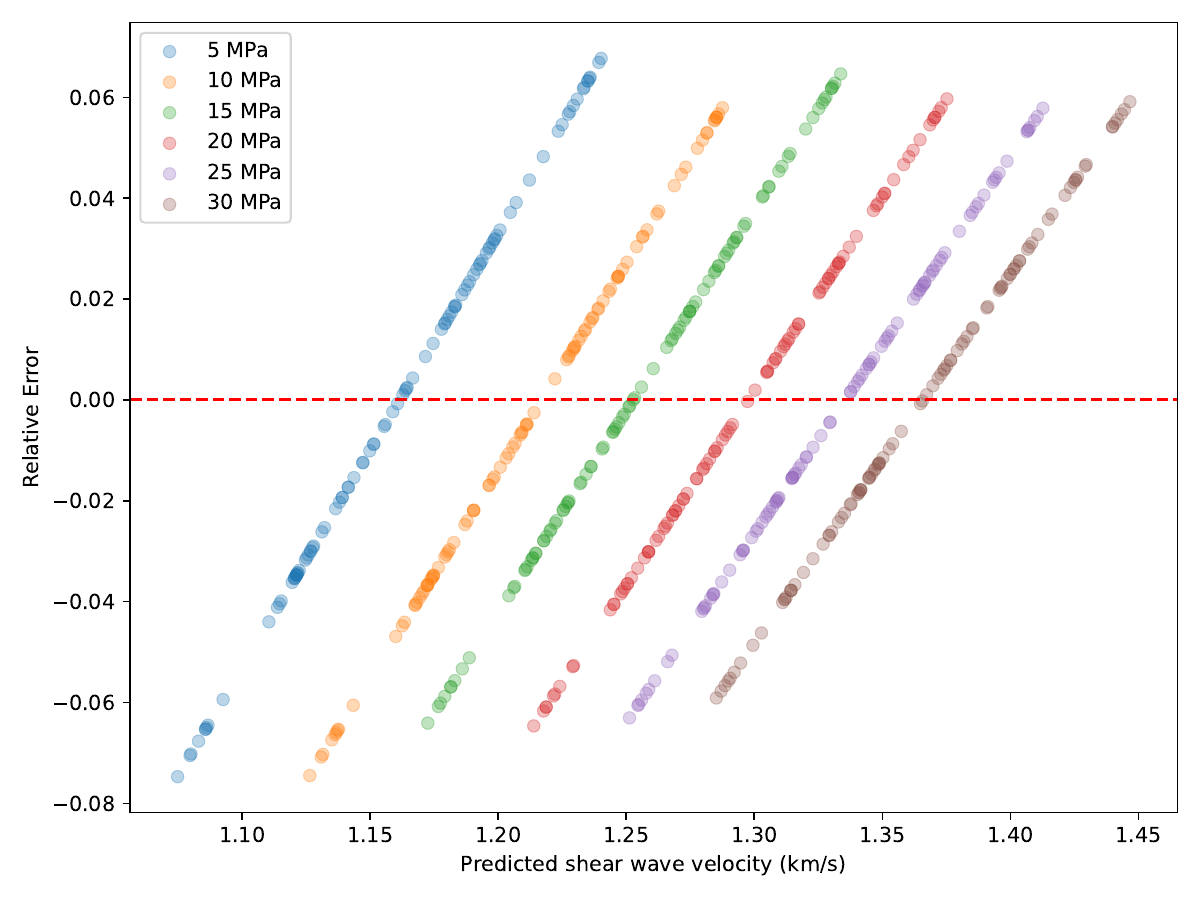}
\caption{Sensitivity analysis on the uncertainty of shear velocity for each effective pressure. (Color online)}
\label{fig3}
\end{figure}

Figs.\ref{fig2} and \ref{fig3} show sensitivity analysis of a $\pm$5\% variation in the optimal coefficients from Eq.(\ref{eq3}) for longitudinal and transverse velocities, respectively. The results indicate that the maximum uncertainties are less than $\pm$5.5\% and $\pm$7.5\% for compressional and shear waves, respectively.

\subsection{Case study}
\begin{figure} [H]
\centering
\includegraphics[width=0.95\textwidth]{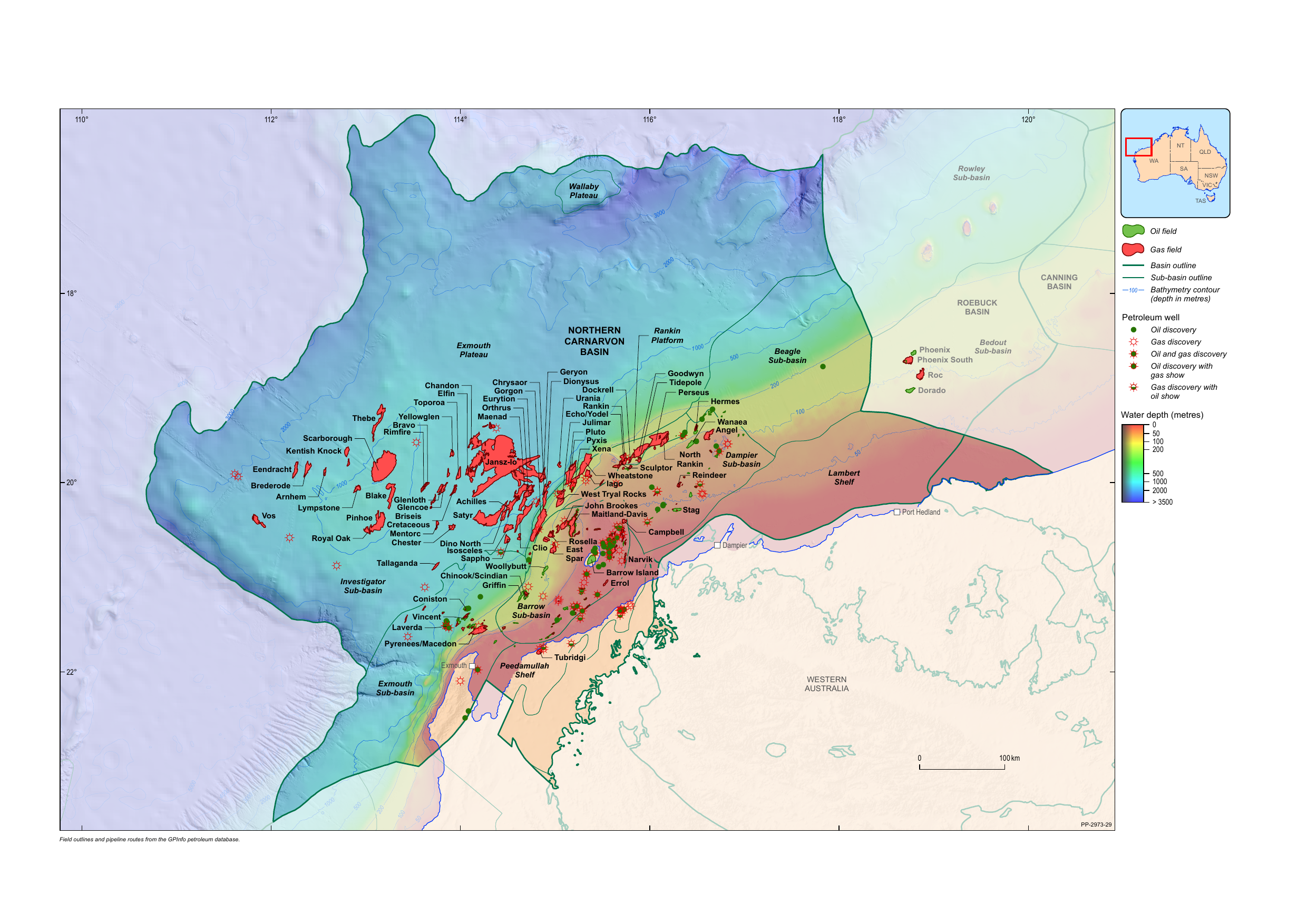}
\caption{Map of the Northern Carnarvon Basin showing bathymetry petroleum well distribution and oil and gas fields. (Color online)}
\label{fig4}
\end{figure}

\begin{figure} [H]
\centering
\includegraphics[width=0.75\textwidth]{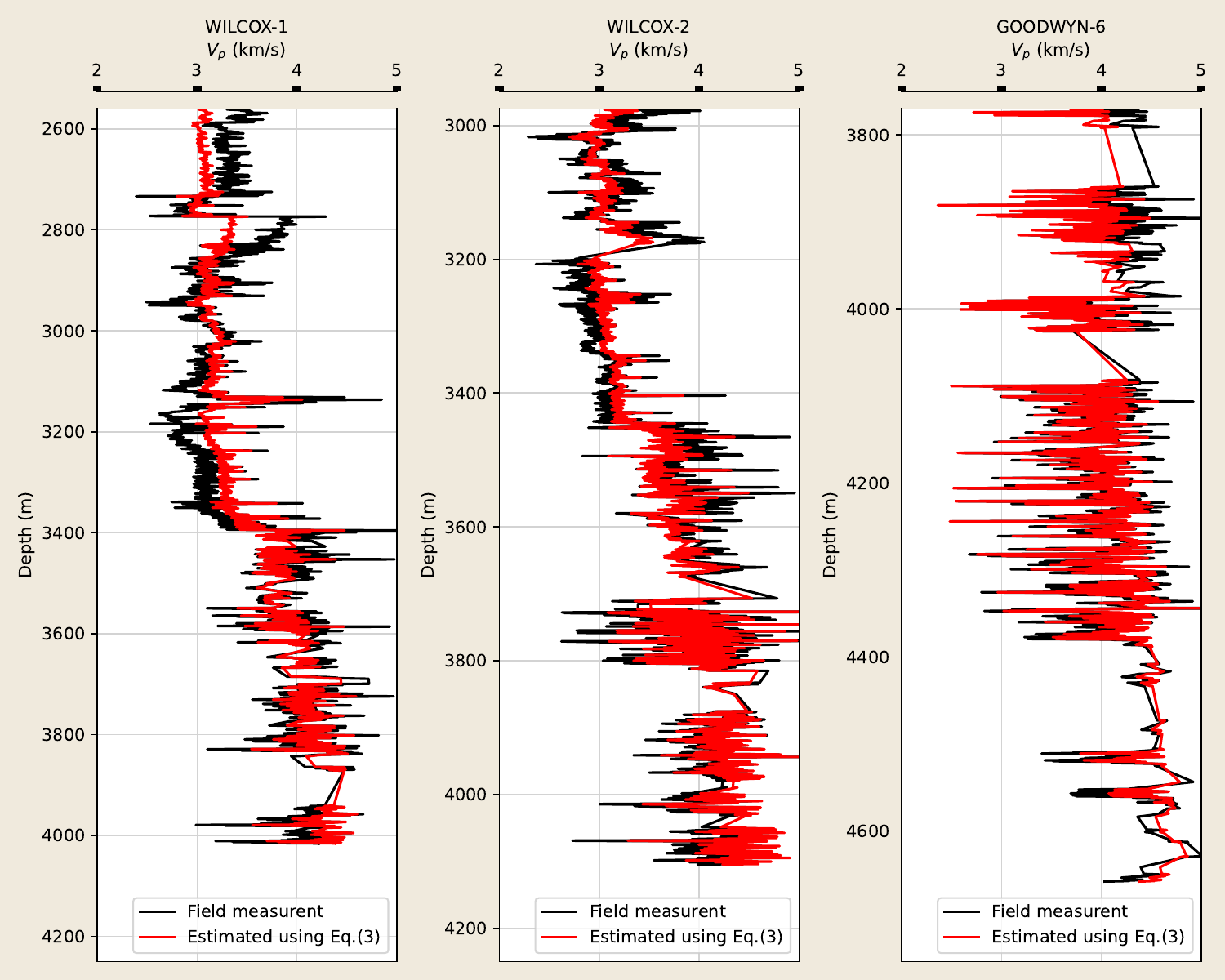}
\caption{Measured and predicted velocity logs for WILCOX-1, WILCOX-2, and GOODWYN-6 wells. The red curve is given by the proposed Eq.(\ref{eq3}). The black curve represents field measurements. (Color online)}
\label{fig5}
\end{figure}

The proposed Eq.(\ref{eq3}) was applied to three wells in the Northern Carnarvon Basin (see Fig.\ref{fig4}): the WILCOX-1 well, WILCOX-2 well, and GOODWYN-6 well. The dataset utilized in this study is sourced from the open-access repository provided by the Department of Mines, Industry Regulation, and Safety, Government of Western Australia. The overburden pressure was determined by density logs, and the Eaton equation was used to predict pore pressure. Effective stress was subsequently calculated as the difference between overburden pressure and pore pressure. Fig.\ref{fig5} presents a comparison between the measured and predicted velocity logs, demonstrating that the estimated values closely align with field measurements for the WILCOX-1, WILCOX-2, and GOODWYN-6 wells. The $\mathrm{R}^2$ for these wells are 0.8246, 0.9152, and 0.8480, respectively.

\section{Conclusions} \label{sec4}
In this study, new relationships were formulated to describe the elastic wave velocity-effective stress correlations for reservoir rocks. The validity of these relationships was verified using published experimental data from 38 core samples. Additionally, their robustness was confirmed through sensitivity analysis. The proposed models were applied to three wells in the Northern Carnarvon Basin, and the results demonstrated that the new elastic wave velocity-effective stress laws yield satisfactory predictions for velocity logs.


\bibliographystyle{elsarticle-harv} 
\bibliography{refs}





% \end{thebibliography}
\end{document}